\documentclass[cite,figures]{epl}
\usepackage{epsfig}
\title{Specific heat and disorder in the mixed state of non-\\
magnetic borocarbides}
\shorttitle{Specific heat and disorder in the...}
\author{D. Lipp\inst{1}\thanks{E-mail: \email{lipp@physik.phy.tu-dresden.de}; 
Fax: +49-351-463-7060} \and M. Schneider\inst{1} \and A. Gladun \inst{1},\\
S.-L. Drechsler\inst{2} \and J. Freudenberger\inst{2}
\and G. Fuchs\inst{2} \and K. Nenkov\inst{2} \and K.-H. M\"uller \inst{2} \and
T. Cichorek\inst{3} \and P. Gegenwart\inst{3} }
\shortauthor{D. Lipp \etal}
\institute{
	\inst{1} Institut f\"ur Tieftemperaturphysik, Technische Universit\"at Dresden, 
D-01062 
Dresden, Germany\\
	\inst{2} Institut f\"ur Festk\"orper- und Werkstofforschung, D-01171 
	Dresden, 
Postfach 270116, Germany\\ 
	\inst{3} Max-Planck-Institut f\"ur Chemische Physik fester Stoffe 
	Dresden,
D-01187 Dresden, Germany
}
\begin{document}

\maketitle

\begin{abstract}
\footnotesize{
The temperature and magnetic field dependence of the specific heat $c_{p}(T,H)$ in the 
superconducting mixed state as well as the upper critical field $H_{c2}(T)$   have been 
measured  for polycrystalline Y$_x$Lu$_{1-x}$Ni$_2$B$_2$C and Y(Ni$_{1-y}$Pt$_y$)$_2$B$_2$C  
samples. The linear-in-$T$ electronic specific heat contribution $\gamma (H) \cdot T$  
exhibits significant deviations from the usual  
linear-in-$H$ law for all $x$ and $y$ the transition metal site ($T$)
%isoelectronically substituted samples,  
resulting in a disorder  dependent negative curvature of $\gamma (H)$. 
 The deviations from that linear behaviour of our  
unsubstituted samples are the largest reported so far for any superconductor. 
 The $H_{c2}(T)$ data point to the quasi-clean limit 
for (Y,Lu)-substitutions and to a transition  to the
quasi-dirty limit for (Ni,Pt)-substitutions.  
The $\gamma(H)$ dependence is discussed in the unitary 
$d$-wave  as well as  in the quasi-clean $s$-wave limits. 
From a consideration of $\gamma (H)$ data only,
$d$-wave pairing
cannot be ruled out.\\}
\end{abstract}

\section{{\bf INTRODUCTION\/}}%%%%%%%%%%%%%%%%%%%%%%%%%%%%%%%%%%%%%%%%%%%%%%%%%%%%%%%%%

The rare earth ($R$) transition metal ($T$) borocarbide family
($R$C)$_n$$T_2$B$_2$ ($R=\mathrm{Y}$, Lu, Sc, Th, La; $T=\mathrm{Ni}$, Pd, Pt;
$n=1$ or 2)  contains  superconductors with relatively high transition temperatures 
$T_c$ up to 23 K \cite{Nagarajan,Cava2}. 
The coexistence of  superconductivity and magnetism for members of this family
where $R$ are magnetic rare earth ions such as Dy, Ho, Er, ... ,
has stimulated numerous studies of their thermodynamic and transport properties 
in the superconducting as well as in the normal state. 
At first glance, most of those results support  
a classification 
of these materials as intermetallic 
phonon mediated 
superconductors with a moderately strong 
coupling strength. However, clean
$R$Ni$_2$B$_2$C samples
 exhibit also
some features unexpected for ordinary $s$-wave  
superconductors.  We emphasize the unusual shape and the
strong disorder dependence of the upper critical 
field $H_{c2}(T)$ and a nearly $T^3$ scaling 
of the electronic specific heat $c_{es}(T)$ in 
the superconducting state compared with
exponential behaviour  for 
ordinary $s$-wave superconductors \cite{drechsler98}. 

According to 
Nohara \textit{et al.}\ \cite{Nohara1} the
isoelectronic $T$-substitution does affect   strongly 
the  field dependence of the linear-in-$T$ 
electronic specific heat contribution $\gamma(H)\cdot 
T$  in the mixed state.  Thus, for an Y(Ni$_{0.8}$Pt$_{0.2})_2$B$_2$C 
single crystal $\gamma (H)\propto H$ has been found, while a square-root law was 
observed for a  pure
YNi$_2$B$_2$C single crystal and for polycrystalline 
LuNi$_2$B$_2$C   \cite{Nohara2}
\begin{equation}
\gamma(H)/\gamma_N \propto \sqrt{H/H_{c2}(0)},
\label{sqrtH}
\end{equation}
where $\gamma_N$ is the 
 Sommerfeld constant in the normal state.
Although  the observed $\gamma (H) \propto \sqrt{H}$-law  for YNi$_2$B$_2$C 
 and LuNi$_2$B$_2$C
 was regarded initially  as  evidence for $d$-wave pairing 
\cite{Nohara2,Maki},  the disorder related transition from a 
$\sqrt{H}$ to a linear-in-$H$
dependence was subsequently used  to rule out  
$d$-wave superconductivity in non-magnetic borocarbides \cite{Nohara1}.  
However, to the best of our knowledge,  systematic investigations of this 
problem in  a broader concentration range for Y(Ni$_{1-y}$Pt$_y$)$_2$B$_2$C 
are lacking. Since isoelectronic substitutions  in the $R$C charge reservoir are expected 
to produce much weaker disorder than those in the  $T$B network we studied also 
the closely related   Y$_x$Lu$_{1-x}$Ni$_2$B$_2$C system 
for the sake of comparison \cite{lipp01}.
 By changing both compositions, $x$ and $y$, deeper insight should be gained on how does the   
disorder affect the field dependence of the specific heat $c_p(T,H)$,   
 the shape and the magnitude of $H_{c2}(T)$,
 as well as   the nature of the pairing state.

\section{{\bf EXPERIMENTAL DETAILS\/}}%%%%%%%%%%%%%%%%%%%%%%%%%%%%%%%%%%%%%%%%%%%%%%%%%

Polycrystalline Y$_x$Lu$_{1-x}$Ni$_2$B$_2$C with $x=0$, 0.25, 0.5, 0.75, 1 and 
Y(Ni$_{1-y}$Pt$_y$)$_2$B$_2$C   
  samples with $y=0$, 0.05, 0.1, 0.15, 0.2, 0.25, 0.5
and 0.75 were prepared by a standard arc melting
technique. Powders of the elements were weighted in the stoichiometric
compositions with a surplus of 10 wt.\% boron to compensate 
losses of boron during arc melting.  The powder was pressed to   
pellets which were melted in argon gas on a water-cooled copper plate
in an arc furnace. To get homogeneous samples, they were turned over and
melted again four times. After the melting procedure the solidified
samples were homogenised at $1100\ ^{\circ}$C for ten days.
The specific heat was measured in the 
 range 4.2 K $\le T \le$ 20 K
and for magnetic fields $\mu _0H \leq$ 8 T using a quasi-adiabatic step heating technique.
The heating pulses were generated by a strain 
gauge heater and $T$ was measured
with a Au-Ge thin film resistor. The temperature dependence of the 
upper critical field
$H_{c2}(T)$ was determined by taking $T_c$ from the jump of $c_p$ in 
the particular field.

\section{{\bf RESULTS AND DISCUSSION\/}}%%%%%%%%%%%%%%%%%%%%%%%%%%%%%%%%%%%%%%%%%%%%%%%%
\begin{figure}[h]
%\vspace{-0.6cm}
\begin{center}
\epsfig{file=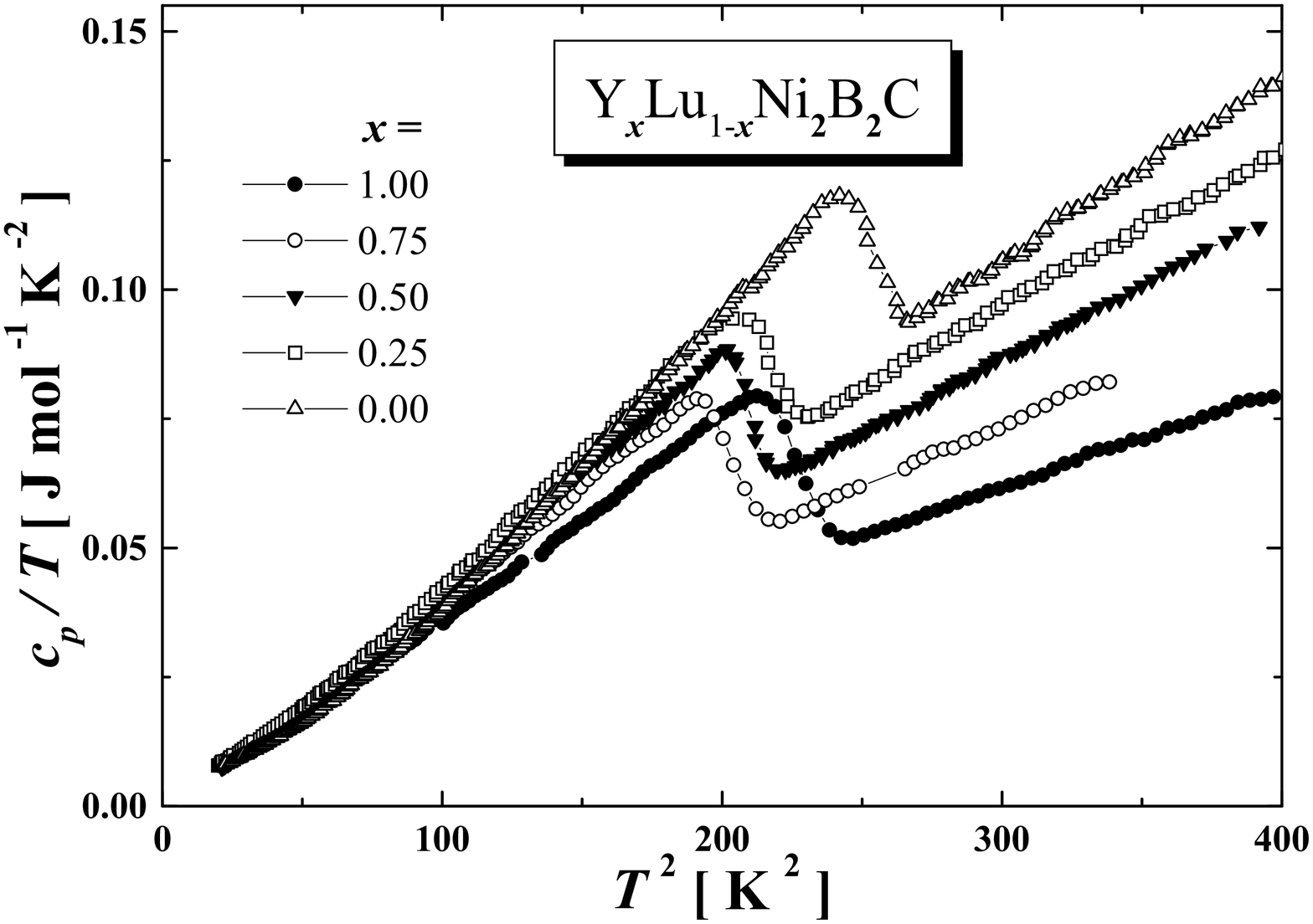,width=7.0cm}
\epsfig{file=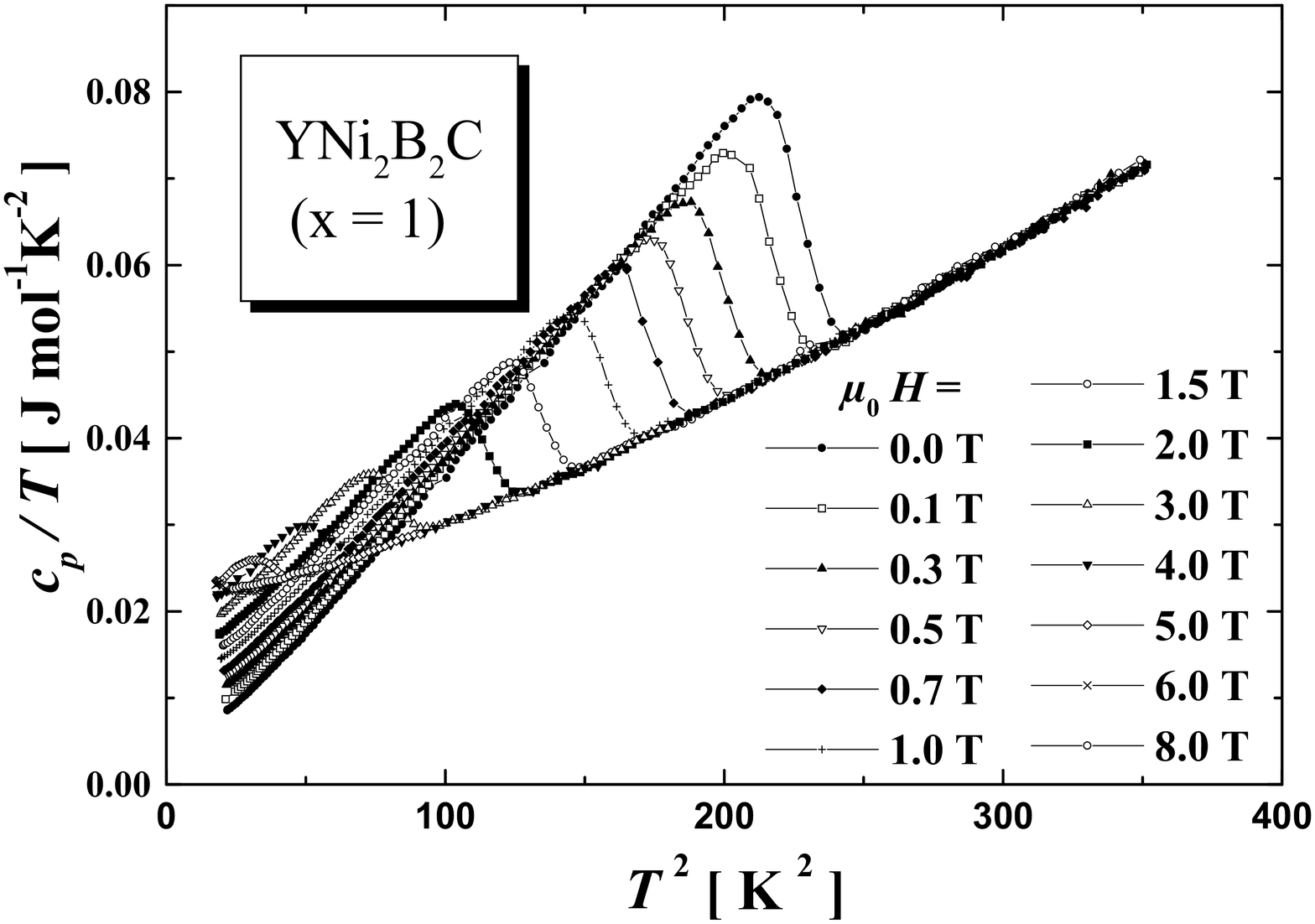,width=7.0cm}
\caption{\footnotesize{Zero magnetic field specific heat  
$c_p(T)/T$ vs $T^2$ of the Y$_x$Lu$_{1-
x}$Ni$_2$B$_2$C series (left panel) and specific heat $c_p(T,H)/T$ vs $T^2$ of 
YNi$_2$B$_2$C for various magnetic 
fields (right).}}
\label{fig1}
\end{center}
\end{figure}

\begin{figure}[h]
\begin{center}
\epsfig{file=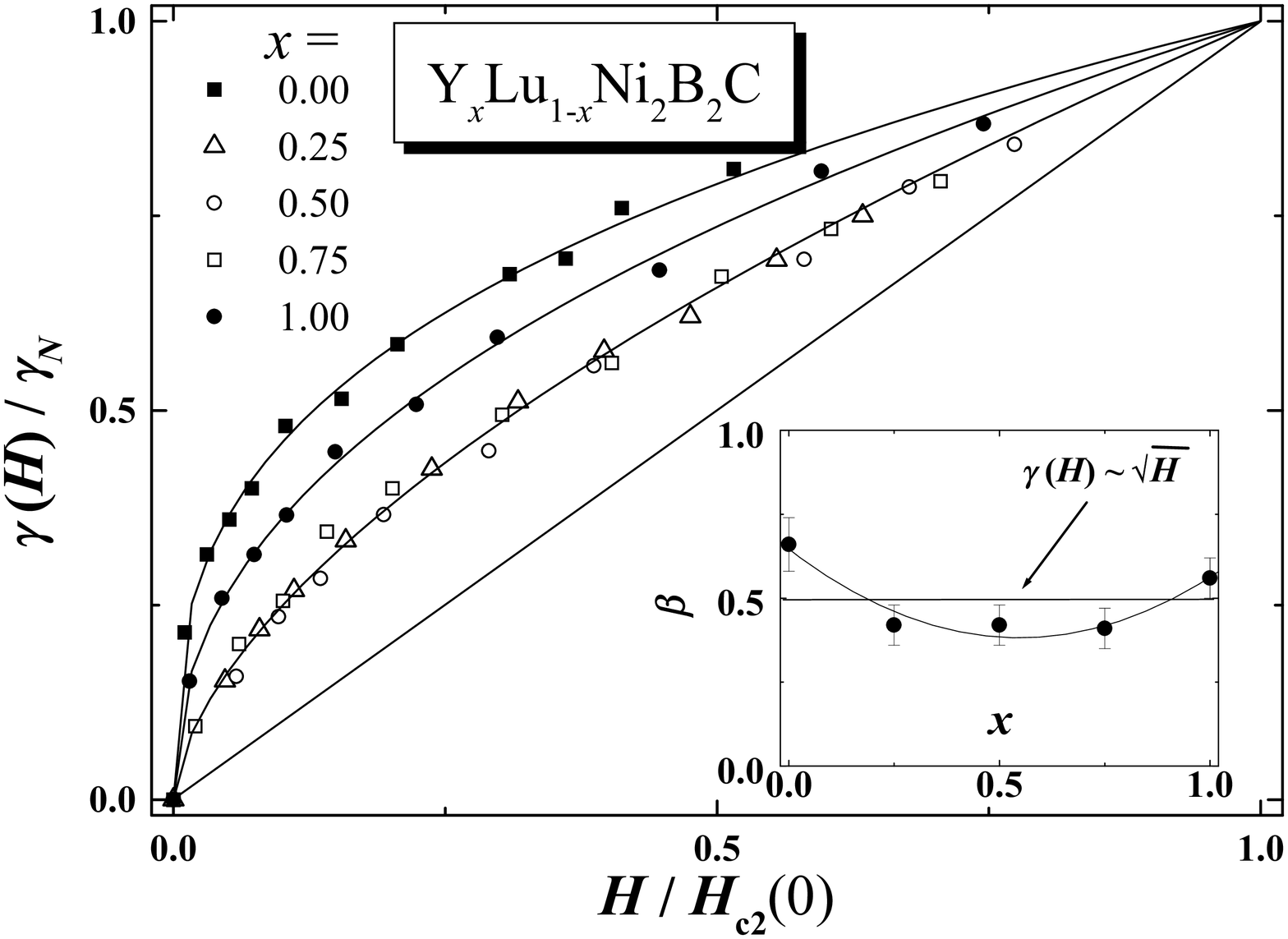,width=7.0cm}
\epsfig{file=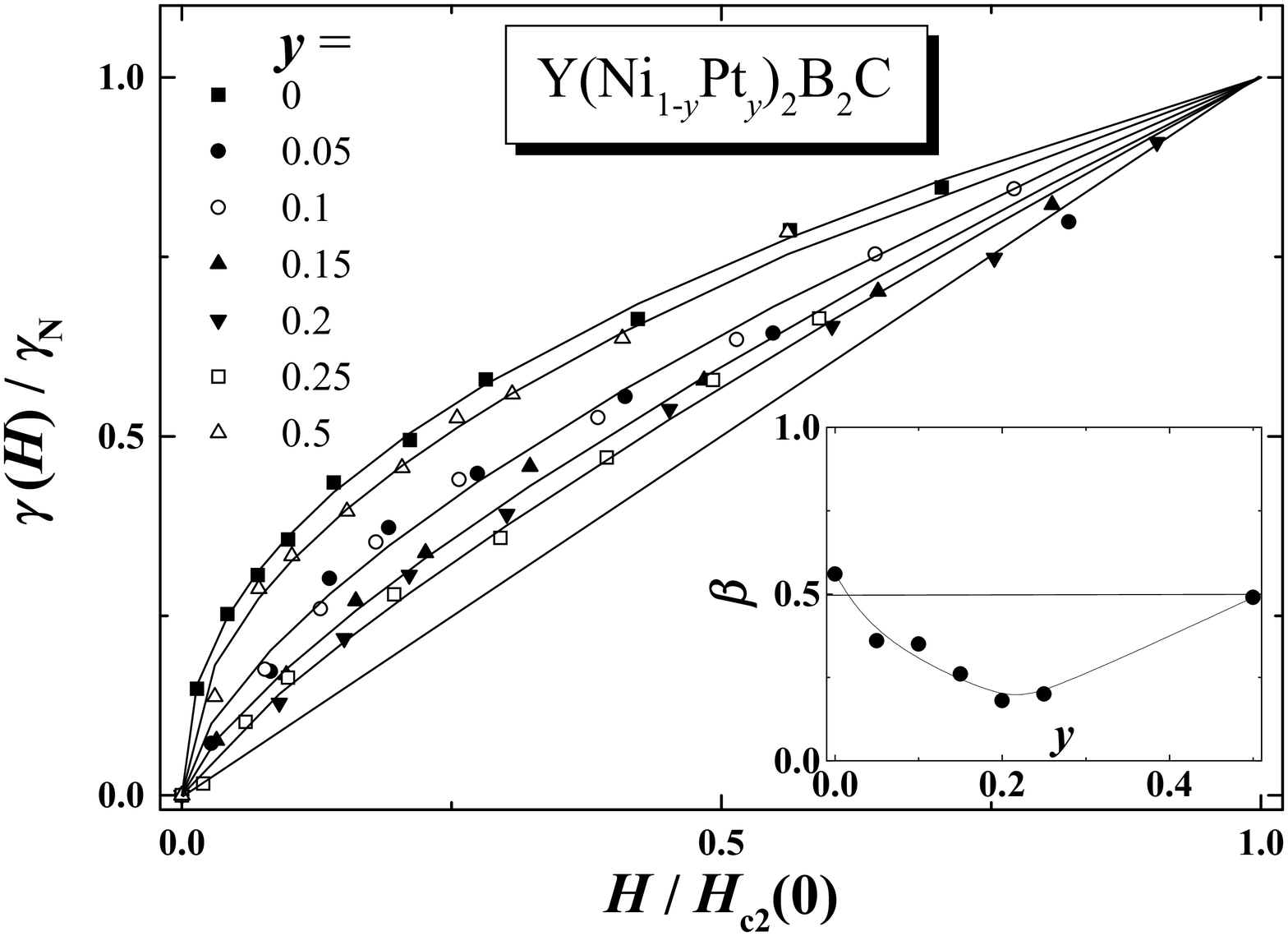,width=7.0cm}
\caption{\footnotesize{Magnetic field dependence of the   
specific heat contribution 
$\gamma (H)$ of the vortex core electrons in the 
mixed state ($H \le H_{c2}$) normalised by $\gamma_N$ with  
 $H_{c2}(0)$ as the upper critical field
(see fig.\ \ref{fig6}) for Y$_x$Lu$_{1-x}$Ni$_2$B$_2$C (left panel) and 
Y(Ni$_{1-y}$Pt$_y$)$_2$B$_2$C (right panel). The lines are fits according 
to eq.\ (\ref{Hbeta}) and the 
straight reference line corresponds to the usual linear-in-$H$ $s$-wave dirty 
limit 
behaviour. 
The insets  show the curvature parameters $\beta (x)$ and  
$\beta (y)$ as defined in 
 eq.\ (\ref{Hbeta}).}}
\label{fig4}
\end{center}
\end{figure}

To illustrate  typical specific heat behaviour, the $c_p/T$ vs $T^2$ data  at $H = 0$ 
of the Y$_x$Lu$_{1-x}$Ni$_2$B$_2$C series  
and the  corresponding 
curves for $\mu _0 H \leq$  8 T of the 
pure Y sample ($x=1$) 
are shown in fig.\ \ref{fig1}.  
Measurements at 8 T were used to analyse 
the    
normal state specific heat 
\mbox{$c_p=\gamma _N T + \beta_D T^3$}, 
where $\gamma _N$ is the Sommerfeld 
constant  and $\beta_D T^3$ is the Debye contribution. 
%%%%%%%%%%%%%%%%%%%%%%%%%%%%%%%%%%%%%%%%%%%%%%%%%%%%%%%%%%%%%%%%%%%%%%%%%%%%%%%%
%%%%%%%%%%%%%%%%%%%%%%%%%%%%%%%%%%%%%%%%%%%%%%%%%%%%%%%%%%%%%%%%%%%%%%%%%%%%%%%%
The  $\gamma_N$ was determined by extrapolating the $c_p/T$ vs 
$T^2$ curves of the high field data in the normal state to $T \to 0$. 
In this way we obtained $\gamma _N=$ 20.4 ($x=0$), 19.0 ($x=0.25$), 
18.3 ($x=0.5$), 18.0 ($x=0.75$) and 20.2 mJ/molK$^2$ ($x=1$) for our 
Y$_x$Lu$_{1-x}$Ni$_2$B$_2$C series in good 
agreement with  the data reported by several groups
\cite{Manalo,fuchs01,drechsler00,Michor1,Movshovich,Hilscher,Carter} and $\gamma_N=20.2$ 
($y=0$), 20.2 ($y=0.05$), 18.4 ($y=0.1$), 16.4 ($y=0.15$), 16.2 ($y=0.2$), 16.9 ($y=0.25$), 
15.3 ($y=0.5$) and 15.0 mJ/molK$^2$ ($y=0.75$)   for the Y(Ni$_{1-y}$Pt$_y$)$_2$B$_2$C 
series. 
To      determine
  $\gamma (H)$, the $c_p/T$ vs $T^2$ 
curves in the particular field have been extrapolated in the same way
to $T \to 0$ from the data in the range 4.2 K $\le T \le 7$ K or up to 
the onset of the transition to the normal state  
(see fig.\ \ref{fig4}).

For all samples $\gamma (H)$ is a
 sublinear function of $H$. At first generalizing eq.\ (\ref{sqrtH}), the 
data were analysed by the expression 
\begin{equation}
\gamma (H)/\gamma_N = [H/H_{c2}(0)]^{1-\beta},
\label{Hbeta}
\end{equation}
where the fitting parameter $\beta$ measures  the sublinearity ({\it i.e.\ } a negative 
curvature) of $\gamma (H)$ and $H_{c2}(0)$ is the field where $\gamma (H)$ 
reaches $\gamma _N$. We obtained $\beta =$ 0.66, 0.42, 0.42, 0.41 and 0.56 ongoing from 
$x=$ 0 to $x=$ 1 for Y$_x$Lu$_{1-x}$Ni$_2$B$_2$C
 and $\beta$ = 0.56, 0.36, 0.35, 0.26, 0.18, 0.20, 0.49 and 0.60 ongoing
from $y=$ 0 to $y=$ 0.75 for Y(Ni$_{1-y}$Pt$_y$)$_2$B$_2$C,  
 with uncertainties of $\Delta \beta / \beta \leq 10\%$
due to the small measured values $\gamma_0 =\gamma (H=0)$ \cite{remark1} and due to 
the procedure used to determine $\gamma (H)$, as mentioned above. 
The dependence of $\beta (x)$ is shown in the inset of \mbox{fig.\ \ref{fig4}} and in
fig.\ \ref{fig6} (left panel). $\beta$ reaches the largest values 
for the bordering cases $x=$ 0 and 1 
and becomes markedly smaller in between. We note that 
our curvatures for LuNi$_2$B$_2$C and YNi$_2$B$_2$C exceed slightly
 the value of $\beta=0.5$  suggested by  eq.\ (\ref{sqrtH}) and  that reported in Refs.\ 
 \cite{Nohara1,Nohara2}. To the best of our knowledge the strong sublinearities 
  for $\gamma (H)$, measured by the exponent $\beta$,  
of the borocarbides under consideration are the largest reported so far
 for any superconductor. The $\beta (y)$ behaviour is depicted in the inset of 
 fig.\ \ref{fig4} and in fig.\ \ref{fig6} (right panel).  Large  
exponents $\beta \approx 0.5$ are observed for $y=$ 0 and $y=$ 0.5, 
whereas $\beta$ is significantly reduced for  Pt-concentrations $y$ inside this 
range and has  a {\it finite} minimum value $\beta=0.18$ at $y=0.2$. This minimum value
 is at variance with the  linear law   for an Y(Ni$_{0.8}$Pt$_{0.2}$)$_2$B$_2$C  
single crystal reported in ref.\ \cite{Nohara1}. We attribute that observation to a stronger 
disorder
compared with our samples.
 Larger Pt-concentrations reveal   even  more pronounced $\beta$ values, 
 \textit{e.g.}\ $y=0.75$; $\beta=0.6$. 
%%%%%%%%%%%%%%%%%%%%%%%%%%%%%%%%%%%%%%%%%%%%%%%%%%%%%%%%%%%%%%%%%%%%%%%%%%%%%%%%

\begin{figure}[h]
\begin{center}
\epsfig{file=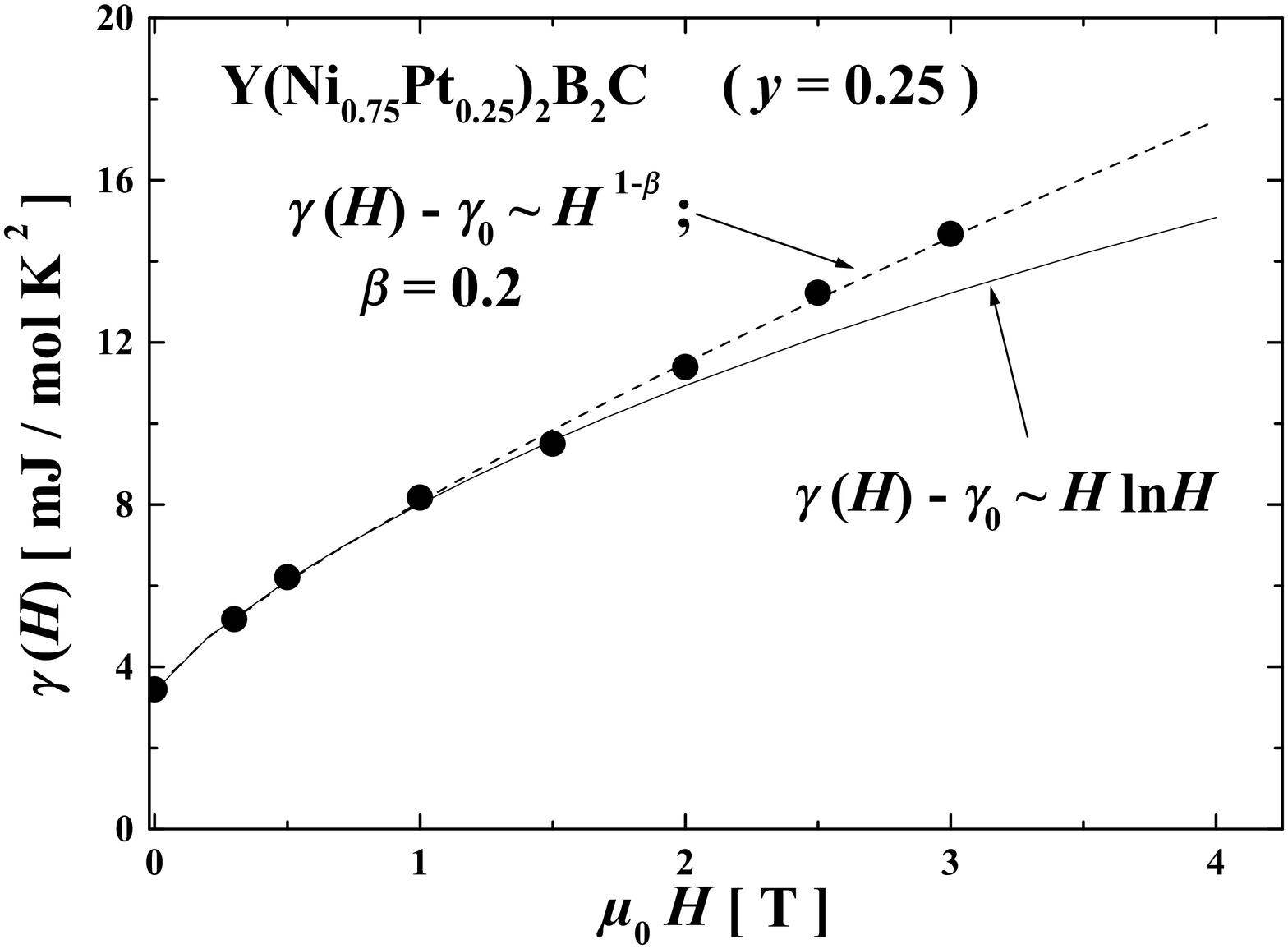,width=7.0cm}
\caption{\footnotesize{Magnetic field dependence of $\gamma(H)$ for 
Y(Ni$_{0.75}$Pt$_{0.25}$)$_2$B$_2$C. The solid line is a fit according to 
eq.\ (\ref{eq.HlnH}).
The dashed line is a fit according to eq.\ (\ref{Hbeta}) with $\beta =0.20$.}}
\label{HlnHy025}
\end{center}
%\vspace{-0.5cm}
\end{figure}

\begin{figure}[hbt]
\begin{center}
\epsfig{file=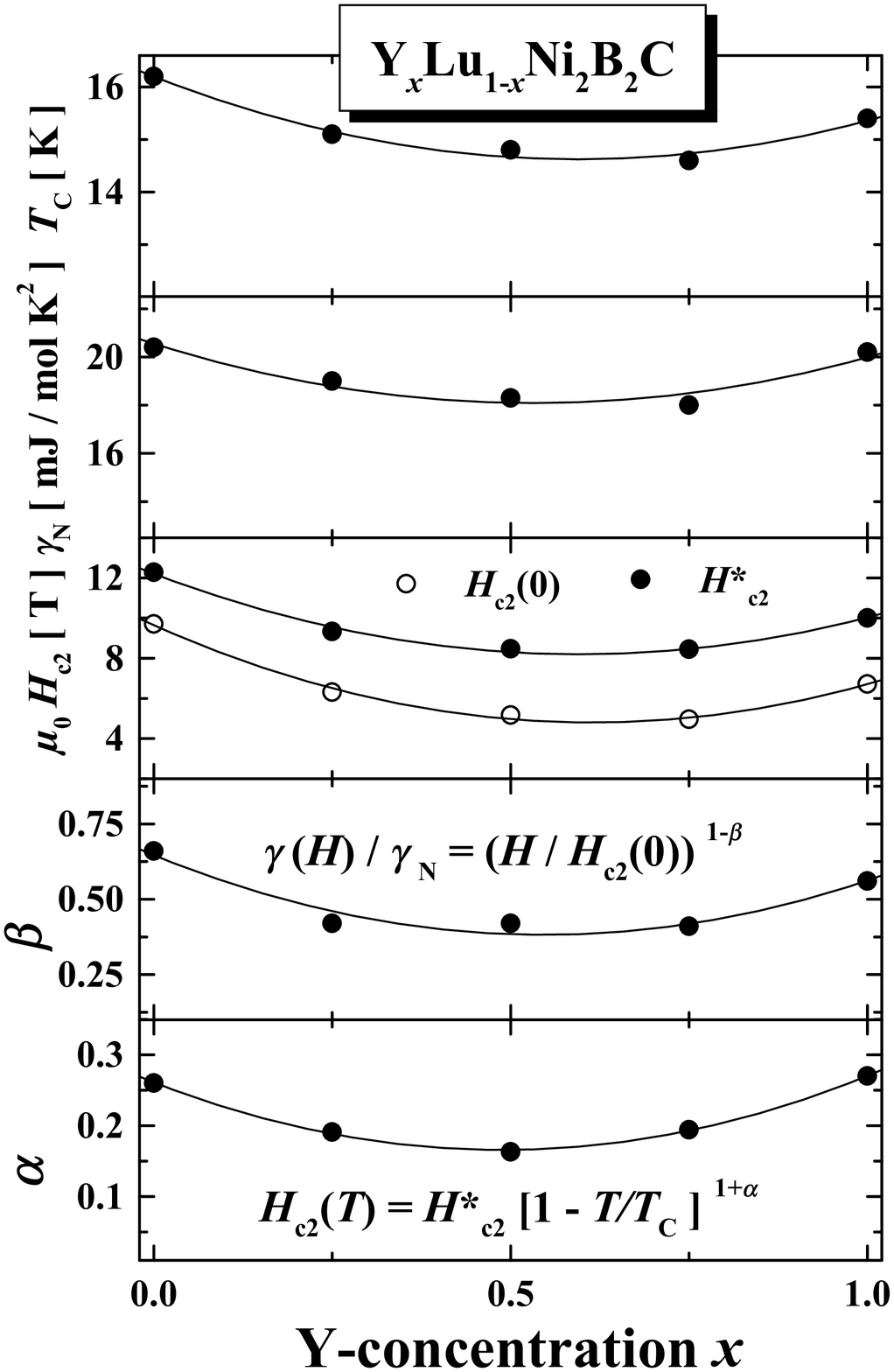,width=7.0cm}
\epsfig{file=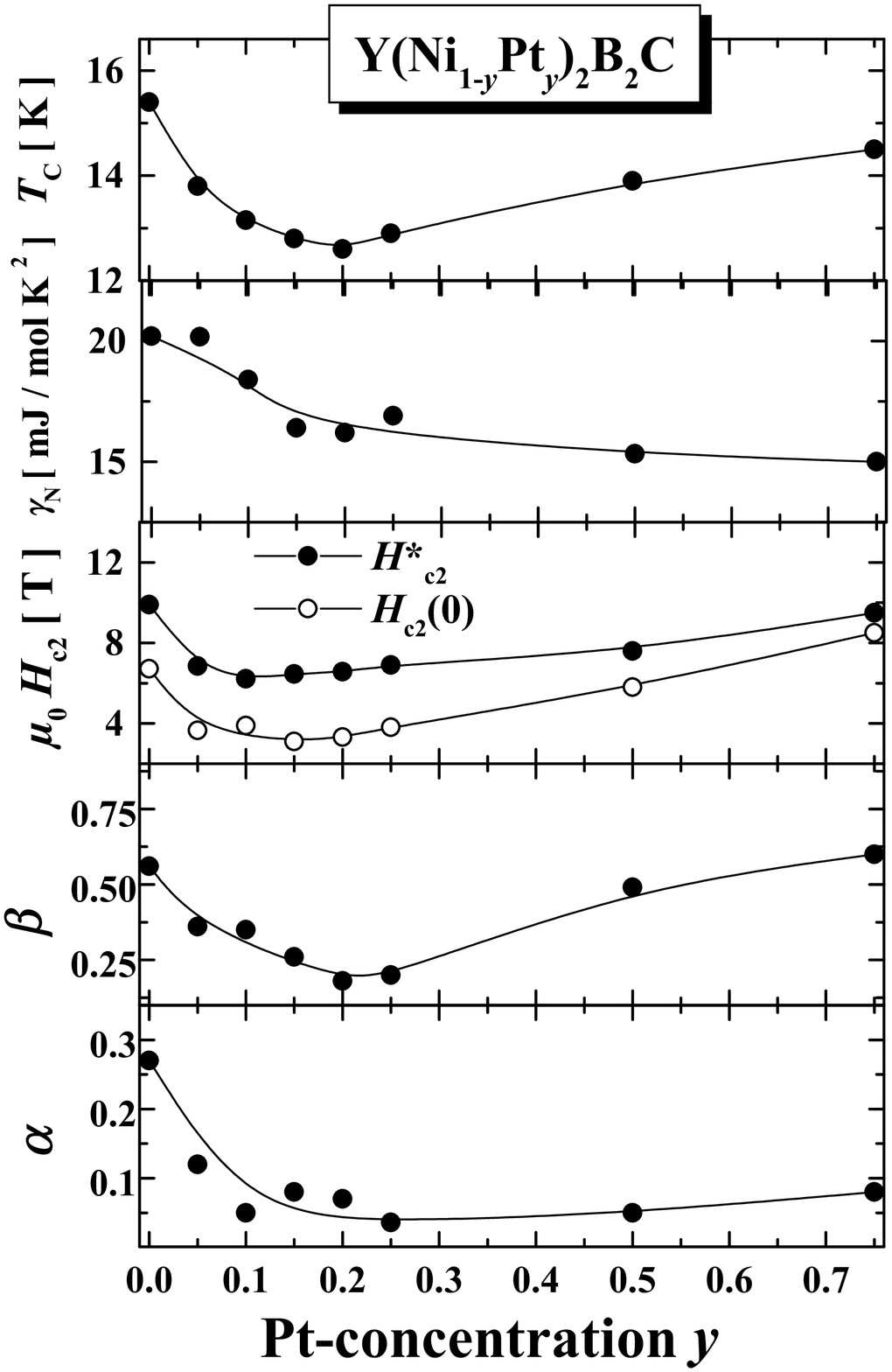,width=7.0cm}
\caption{\footnotesize{Composition dependence of the transition temperature $T_c$ derived 
from the onset 
of the jump of $c_p(T)$ (upper panels), the 
Sommerfeld constant $\gamma _N$ (second panels), 
the upper bound for the upper critical
 field $H^*_{c2}$ according to eq.\ (\ref{Hc2}) and $H_{c2}(0)$ 
according to eq.\ (\ref{Hbeta}) 
(third panels, see text for more details),
the specific heat curvature exponent  $\beta$ of 
$\gamma(H)$ according to eq.\ (\ref{Hbeta}) (fourth panels), and the 
curvature exponent $\alpha$ of the upper critical field 
$H_{c2}$ according to eq.\ (\ref{Hc2}) (lower 
panels) determined for Y$_x$Lu$_{1-x}$Ni$_2$B$_2$C (left ) and for 
Y(Ni$_{1-y}$Pt$_y$)$_2$B$_2$C (right). The lines are guides to the eye.}}
\label{fig6}
\end{center}
%\vspace{-0.5cm}
\end{figure}

%%%%%%%%%%%%%%%%%%%%%%%%%%%%%%%%%%%%%%%%%%%%%%%%%%%%%%%%%%%%%%%%%%%%%%%%%%%%%%%%
The observed $\gamma (H) \propto H^{1-\beta}$-law with $\beta \approx $ 0.5
raises the question 
whether an 
unconventional pairing mechanism is responsible for this
peculiarity since according to Ref.\ \cite{Volovik}
$\gamma (H) \propto \sqrt{H}$ is a signature 
for a nodal order parameter with $d$-wave symmetry (a somewhat
larger value
$\beta =0.59$ has been found in Ref.\ \cite{ichioka99}) while
 $\gamma (H) 
\propto \ H$ describes  an isotropic   $s$-wave order parameter. 
According to Refs.\ \cite{Kuebert2,barash97} Volovik's clean limit $d$-wave 
approach can be generalized  to describe also strong impurity scattering.
 Then at low 
magnetic fields $H \ll H_{c2}(0)$ the specific heat coefficient $\gamma (H)$ follows  
a $H\ln H$ dependence:
\begin{equation}
\gamma(H) = \gamma_0 + \gamma_N D 
\left(\frac{H}{H_{c2}(0)}\right)\ln \left[\frac{\pi}{2a^2}
\left(\frac{H_{c2}(0)}{H}\right)\right],
\label{eq.HlnH}
\end{equation}
where $a$ and $D$ are constants. 
Such a   behaviour was observed for 
various disordered high-$T_c$ cuprates and considered as evidence for
$d$-wave superconductivity in the unitary scattering limit \cite{nohara3,nohara4}.
At the same time its applicability to   non-magnetic 
borocarbides under consideration was disclaimed\cite{nohara3}. However, some 
 of our data can be described equally well by eq.\ (\ref{eq.HlnH}) for 
$H/H_{c2}(0)\leq 0.3$ as well as by eq.\ (\ref{Hbeta}) using
intermediate values for $\beta$ (0.2 to 0.35). This is shown in
  fig.\ \ref{HlnHy025}: obviously,  
the $H$ln$H$ behaviour is not very distinct from 
the power law at low fields $\mu _0H \leq 1.5$ T. At higher 
fields the $H$ln$H$ dependence may deviate since it was derived 
for low fields only \cite{Kuebert2}). 
For $y=$ 0.25 and 0.5 we observed residual values $\gamma_0=3.4$ and 3.3 mJ/molK$^2$,
respectively, which are substracted in fig.\ \ref{fig4} \cite{remark1}. 
The existence of a non-negligible $\gamma_0$ is a feature predicted for a $d$-wave 
order parameter
in the unitary limit \cite{Preosti}. 
Hence, $d$-wave pairing cannot be ruled out in non-magnetic borocarbides
by considering  $\gamma (H)$ data only.    
%%%%%%%%%%%%%%%%%%%%%%%%%%%%%%%%%%%%%%%%%%%%%%%%%%%%%%%%%%%%%%%%%%%%%%%%%%%%%%%%%%%%%%% 
%%%%%%%%%%%%%%%%%%%%%%%%%%%%%%%%%%%%%%%%%%%%%%%%%%%%%%%%%%%%%%%%%%%%%%%%%%%%%%%%%%%%%%%
While the deviation from the linearity of $\gamma (H)$ is frequently 
ascribed to a shrinking of the vortex cores with magnetic field and to vortex core 
interactions 
\cite{Nohara1,SonierPRL82_4914,IchiokaPRB59_8902}, recent 
investigations support the assumption of delocalized quasiparticle states 
around the vortex core to be responsible for this feature, 
similar as in  $d$-wave superconductors \cite{IzawaPRL86_1327}.
However, there are  several conventional, but anisotropic $s$-wave, 
superconductors which also exhibit  deviations from the  
$\gamma (H) \propto H$-law in the clean limit,  \textit{e.g.}\  
V$_3$Si \cite{Ramirez}, NbSe$_2$ \cite{Nohara1} 
($\beta=0.33$),
 and CeRu$_2$ \cite{ichioka99,Hedo}. Remarkably, 
a sublinear $\gamma (H)$ behaviour has been reported also for the novel 
``medium-$T_c$'' superconductor MgB$_2$ \cite{Wang_cond-mat/0103181}.
In this general context recent ultrahigh-resolution photoemission spectroscopy
measurements suggest that a highly anisotropic gap might be responsible
for the mentioned above peculiarities in clean Ni borocarbides \cite{yokoya2000}.
By introducing disorder ($y=0.2$) a complete isotropization of the
gap  was observed. Calculations of the density of states (DOS) at the Fermi level,  
$N(0)$, in the mixed state with interacting vortices 
by Ichioka \textit{et al.}\ \cite{ichioka99}
 revealed  
 a $H^{0.67}$ ($\beta=0.33$) dependence of $\gamma (H)$ for anisotropic 
$s$-wave superconductors. 

In this context it is noteworthy that 
 our low-temperature $c_p$ data on YNi$_2$B$_2$C ($x=1$) point to a possible 
 exponential behaviour of $c_{es}(T)$ below 1/6 $T_c$ with 
 $\Delta (0)/k_B T_c \approx 0.5$. This supports an effective
multiband $s$-wave picture like that proposed by Shulga \textit{et al.}\ \cite{Shulga}. 
Noteworthy, a similar 
value for $\Delta (0)$ was observed in microwave measurements 
\cite{JacobsPRB52_R7022}. 

Like $\gamma(H)$, the upper critical field $H_{c2}(T)$ can be described also by a simple 
scaling law
\cite{Freudenberger1}
\begin{center}
\begin{equation}
H_{c2}(T)=H^*_{c2}(1-T/T_c)^{1+\alpha},\quad \textrm{valid for} \quad 0.3 
\leq T/T_c .
\label{Hc2}
\end{equation}
\end{center}
Our values of the upper critical field $H_{c2}(0) \approx 0.9 H^*_{c2}$ are reduced 
due to R-site substitution.
 A similar behaviour was found for the pronounced positive curvature of 
$H_{c2}(T)$ 
near $T_c$, which
is measured by the exponent
$\alpha$ in eq.\ (\ref{Hc2}),
in contrast to the opposite statement of a nearly constant curvature \cite{Manalo}. 
The unusual positive curvature of $H_{c2}(T)$ near $T_c$ observed here can be 
explained for superconductors in the clean limit
by  a significant dispersion
of the Fermi velocities using \textit{e.g.}\   an effective  
two-band model \cite{Shulga}.
%%%%%%%%%%%%%%%%%%%%%%%%%%%%%%%%%%%%%%%%%%%%%%%%%%%%%%%%%%%%%%%%%%%%%%%%%%%%%%%%
%\vspace{0.0cm} 
%%%%%%%%%%%%%%%%%%%%%%%%%%%%%%%%%%%%%%%%%%%%%%%%%%%%%%%%%%%%%%%%%%%%%%%%%%%%%%%%%
\noindent
$T_c$ and $\gamma _N$ are  reduced  to a smaller extent, which has been ascribed to
a slight reduction of the electron-phonon coupling constant $\lambda$ 
at intermediate $x$\cite{drechsler99a,freudenberger99}.
For  $T_c$ a dip near 
$x=0.7$ is observed, in accordance with refs.\ \cite{Manalo,Freudenberger1}
 ($T_c \approx 14.6$ K at $x=0.75$).  The dirty limit region
is not reached (which would be represented by vanishing $\alpha$ and increasing
$H_{c2}(0)$ with increasing disorder \cite{fuchs01,drechsler00}

In the case of Pt-substitutions in the investigated range $0< y < 0.75$, 
the values of $T_c$, $\beta$ and $H_{c2}(0)$ are reduced, too.
As for  R-substitutions, those superconducting properties exhibit minima at 
intermediate composition
while the Sommerfeld constant $\gamma_N$ and the curvature parameter 
$\alpha$ of $H_{c2}(T)$ depend monotonically on $y$. For $y < 0.2$ a strong
decrease of $\alpha$ with increasing $y$ is observed, 
but 
for $y > 0.2$ an increase of $\alpha$ does not occure (see fig.\ \ref{fig6}). 
This  behaviour of $H_{c2}$ suggests that the quasi-dirty limit has been 
reached at about $y\approx0.2$ since 
$H_{c2}(0)$ increases above $y\approx 0.2$, while the curvature of $H_{c2}$ 
measured by $\alpha$ remains   strongly reduced. 
The results obtained for  Y(Ni$_{1-y}$Pt$_y$)$_2$B$_2$C show that 
the deviations from the linearity  of $\gamma (H)$ measured by $\beta$ are 
not  correlated  with the field exponent $\alpha$.
 While $\alpha$ almost vanishes, 
$\beta$ does increase for $y > 0.2$. Thus, here 
the behaviour of the specific heat in the 
vortex state even in the quasi-dirty limit remains rather complex.
  In this context a similar behaviour for (Nb,Ti) alloys which are 
apparently strongly disordered superconductors is noteworthy: on one hand,
a strong sublinear $\gamma(H)\propto H^{0.5}$-dependence is observed, and on the 
other hand,  a standard parabolic Werthammer-Helfand-Hohenberg 
temperature dependence of $H_{c2}(T)$ occures, {\it i.e.\ }
 a negative curvature near $T_c$ is present \cite{lipp2001}.    

To summarize, we have shown that the $\gamma (H)$ curves of the pure specimens ($x=$ 0; 1)
of Y$_x$Lu$_{1-x}$Ni$_2$B$_2$C exhibit the strongest sublinear behaviour 
reported for 
superconductors. 
Weak disorder effects caused by isoelectronic substitutions of Lu by 
Y yield a reduction of the $\gamma (H)$-non-linearity without
reaching the standard linear behaviour.  
Similar moderate suppressions of characteristic features which are typical for 
the quasi-clean limit have been found for the upper critical field $H_{c2}(0)$, 
the  curvature exponent $\alpha$,  $\gamma _N$ and  $T_c$. 
Stronger disorder effects are caused by isoelectronic substitutions of Ni by Pt.
 From the behaviour of $H_{c2}(T)$ it is 
 deduced that a transition from clean to quasi-dirty limit occures 
caused by isoelectronical substitutions at the $T$-site.
 The quasi-dirty limit is deduced from the nearly vanishing curvature of 
 $H_{c2}(T)$ for $y\geq 0.2$.
 At the same time there the sublinearity of $\gamma (H)$ remains and does even 
 increase. Hence, a simple monotonical relationship between $\alpha$ and $\beta$, 
  as one might expect by considering 
the results on Y$_x$Lu$_{1-x}$Ni$_2$B$_2$C only, 
 does not hold in the quasi-dirty limit. 
In the case of intermediate deviations from the linearity of 
$\gamma (H)$ 
($\beta = 0.2 - 0.35$)
our results on specific
heat at low magnetic fields are discussed in the context of a dirty $d$-wave 
model on the one hand and
within the framework of the conventional $s$-wave picture in the quasi-clean 
limit on the other hand. At low fields the $H$ln$H$ dependence 
of $\gamma (H)$ predicted for $d$-wave pairing in the dirty (unitary) limit
is not very distinct from the $H^{1-\beta}$ behaviour 
which favoures $s$-wave superconductivity in the quasi-clean limit. 
Thus, results on $\gamma (H)$ 
only are insufficient to rule out  unconventional pairing  in 
borocarbide superconductors. But 
from the few available $c_p$ data points at  low-temperature  and $H=0$ a small gap of 
$\Delta (0)/k_B T_c \approx 0.5$ might be
 derived  for a clean 
YNi$_2$B$_2$C sample. 
This reveals additional 
support for an anisotropic $s$-wave pairing state in non-magnetic 
borocarbides.
%\vspace{-5mm}
\acknowledgments
This work has been supported by the SFB 463 and the DFG. 
We  acknowledge discussions with S.\ Shulga, H.\ Rosner, H.\ Michor, 
M.\ Nohara, K.\ Maki, H.\ Takagi, and D.G.\ Naugle as well as R.\ Botha 
for a critical reading of the manuscript.


\begin{thebibliography}{99}
\bibitem{remark1} For all our other samples small residual values 
$\gamma _0 \leq 1.5$mJ/molK$^2$ are observed.
% which can be ascribed either to
%the existence of a small content of a non-superconducting phase 
%or to the feature of the $d$-wave state in the unitary limit.
These values are substracted in fig.\ \ref{fig4}, too.
\end{thebibliography}
\end{document}